 \documentclass[pra,showpacs,aps]{revtex4}

\usepackage{dcolumn}
\usepackage{graphicx}
\begin{document}

  \newcommand{\eref}[1]{(\ref{#1})}
  \newcommand{\Eq}[1]{eq.~(\ref{#1})}
  \newcommand{\Eqs}[1]{eqs.~(\ref{#1})}
  \newcommand{\Table}[1]{Table~\ref{#1}}
  \newcommand{\Tables}[1]{Tables~\ref{#1}}
  \newcommand{\Fig}[1]{Fig.~\ref{#1}}
  \newcommand{\cm}{\mbox{cm}\ensuremath{^{-1}}}
\title
{Accounting for Breit interaction in actinide and superheavy element compounds:
    1.~General remarks.}

\author{A.\ N.\ Petrov}
\email{anpetrov@pnpi.spb.ru}
\homepage{http://www.qchem.pnpi.spb.ru}
\affiliation{Petersburg Nuclear Physics Institute,
              Gatchina, Leningrad district 188300, Russia}
\author{N.\ S.\ Mosyagin}
\affiliation{Petersburg Nuclear Physics Institute,
              Gatchina, Leningrad district 188300, Russia}
\author{A.\ V.\ Titov}
\affiliation{Petersburg Nuclear Physics Institute,
              Gatchina, Leningrad district 188300, Russia}
\author{I.\ I.\ Tupitsyn}
\affiliation{Physics Department, St.-Petersburg State University, 
Starii Petergoff, St.-Petersburg 198904, Russia}
\date{\today}

\begin{abstract} The incorporation of the Breit interaction in atomic and
 molecular calculations is discussed in the framework of four-component
 all-electron and two-component relativistic effective core potential (RECP)
 formalisms.  Contributions of the Breit interaction between different core and
 valence shells are studied in the Dirac-Fock approximation for the uranium,
 plutonium, ekathallium ($Z$=113), ekalead ($Z$=114) and other heavy atoms.
 It is shown that the two-electron Breit effects between the valence electrons
 can be neglected for ``chemical accuracy'' (1 kcal/mol) of calculation of
 spectroscopic properties of systems containing superheavy elements and
 actinides whereas large core-core and core-valence Breit contributions can be
 efficiently described by one-electron RECP operators.  Different versions of
 the generalized RECPs with the Breit interaction taken into account are
 constructed for uranium and plutonium and tested in comparison with the
 corresponding all-electron four-component calculations.
\end{abstract}

\pacs{31.10.+z, 31.15.Fx, 31.15.Ne, 31.30.Jv}

\maketitle

\section*{Introduction.}

 The Dirac Hamiltonian which accounts for relativistic effects for an
 electron in an external potential $v$ can be written in the form

\begin{equation}
     h^{\rm D}=c{\vec \alpha}{\cdot}{\vec p} + \beta mc^2 + v\ ,
 \label{one}
\end{equation}
 where $\alpha$ and $\beta$ are Dirac matrices, $c$ is velocity of light. The
 generalization of this equation to the many-electron case is not as
 straightforward as is in the nonrelativistic case.  For few-electron systems,
 the quantum electrodynamics (QED) calculations can be usually performed
 \cite{Shabaev:02a,Labzowsky:02b}.  However, the perturbative QED approach is
 not applicable in practice to many-electron systems such as heavy atoms and
 their compounds when correlation effects are also very important.  Instead,
 the Hamiltonian formalism with some approximations for one- and two-electron
 parts is employed.  For many-electron systems, the effective Hamiltonian (in
 which relatively small irreducible three-electron etc.\ interactions
 are neglected) is usually written as

\begin{equation}
    {\bf H}\ =\ \sum_i h^D_i + \sum_{i>j} G_{ij}\ ,
 \label{Heff1}
\end{equation}
 where $G_{ij}$ is an effective operator which describes interaction between
 electrons $i$ and $j$.  It can be most naturally derived from QED.
 In the first QED order, this operator describes interelectronic exchange by 
 a single virtual photon. The problem is that the $G_{ij}$ operator is gauge 
 dependent in this approximation, e.g., in Feynman gauge 

\begin{equation}
   G_{ij}^F(\omega_{ij}) =
         (1-{\vec \alpha}_i{\cdot}{\vec \alpha}_j)
          \frac{\cos (\omega_{ij} r_{ij})}{r_{ij}}
   \ ,
 \label{BijwF}
\end{equation}
 where $\omega_{ij}$ is the frequency of the exchanged photon divided by the
 velocity of light, and in Coulomb gauge 
\begin{equation}
   G_{ij}^C(\omega_{ij}) =
        1/r_{ij} -({\vec \alpha}_i{\cdot}{\vec \alpha}_j)
          \frac{\cos (\omega_{ij} r_{ij})}{r_{ij}}
        + \left[{\vec \alpha}_i{\cdot}{\vec \nabla}_i,
          \left[{\vec \alpha}_j{\cdot}{\vec \nabla}_j,
       \frac{\cos (\omega_{ij} r_{ij}){-}1}{\omega_{ij}^2 r_{ij}}\right]\right]
   \ .
 \label{BijwC}
\end{equation}
 As is shown in \cite{Hata:84}, operators \eref{BijwF} and \eref{BijwC} give
 equal mean values with one-configuration wave function
constructed of one-particle solutions of \Eq{one} if $v({\vec r})$ is a local
potential.  However, it is demonstrated both theoretically and computationally
in \cite{Gorceix:87,Gorceix:88} that the result is gauge-dependent
 in a general case.  Then, it was clarified in \cite{Lindroth:89b} that the
 gauge dependence remains if one goes beyond the no-virtual-pair approximation
 as well.  It was noted in \cite{Lindgren:90} that when solving the eigenvalue
 problem with Hamiltonian~\eref{Heff1}, only so called ladder diagrams
 of a multi-photon exchange are taken into account.  When two-photon crossed
 diagram is also considered, the two gauges give the same result in the
 leading relativistic order, ${\cal O}(\alpha^2)$\,, where
 $\alpha{\approx}${1}{/}{137} is the fine structure constant.
 Because the crossed diagram does not contribute in this order in the Coulomb
 gauge \cite{Lindgren:90}, the electron-electron interaction in
 form~(\ref{BijwC}) is
 usually more appropriate for electronic structure calculations based on
 eq.~(\ref{Heff1}).

 In this paper we study the Breit interaction in the Dirac-Fock approximation.
 For the reason discussed below, we will further use the low-frequency limit
 of eq.~(\ref{BijwC})

\begin{equation}
   G_{ij}^C(0) =
        1/r_{ij}
-{\vec \alpha}_i{\cdot}{\vec \alpha}_j/r_{ij}\ {+}
                 \frac{1}{2}\left[{\vec \alpha}_i{\cdot}{\vec \alpha}_j {-}
                ({\vec \alpha}_i{\cdot}{\vec r}_{ij})
                ({\vec \alpha}_j{\cdot}{\vec r}_{ij})/r_{ij}^2\right]/r_{ij}\
   \ .
 \label{Bij0C}
\end{equation}
 The first term in eq.~(\ref{Bij0C}) describes the instantaneous Coulomb
 interaction between electrons and the following two terms correspond to the
 original Breit interaction (BI).  The second term in \Eq{Bij0C}
 describes instantaneous magnetic or {Gaunt} interaction which is usually
 a dominant part of BI for an atomic system and the third term describes
 {classical retardation} of the Coulomb interaction between electrons.
 Operator \eref{Heff1} with electron-electron interaction \eref{Bij0C} is
 called Dirac-Coulomb-Breit (DCB) Hamiltonian.  If the retardation part or both
 the retardation and magnetic parts of BI are neglected, one goes to the
 Dirac-Coulomb-Gaunt (DCG) or Dirac-Coulomb (DC) Hamiltonians, correspondingly.

 The terms of order of ${\cal O}(\omega_{ij}^2r_{ij})$\,, which represent
 higher order retardation, are omitted in low-frequency approximation
 \eref{Bij0C} for electron-electron interaction \eref{BijwC}.
 This approximation is not valid for BI between negative (positronic)
 and positive (electronic) energy states, when $\omega_{ij}c \approx 2mc^2$.
 When low-frequency approximation \eref{Bij0C} is applied to derive the
 positive and negative energy one-particle states as solutions of the
 Dirac-Fock-Breit equations, the errors for the energies are of order of
 ${\cal O}(\alpha^4)$ as is shown in~\cite{Lindroth:89c}.
 A serious disadvantage of the high-order retardation terms is that they lead
 to the non-Hermitian exchange part of the Dirac-Fock potential
 \cite{Lindroth:89c}.  It is pointed out in \cite{Feynman:49} that inclusion of
 higher order retardation effects in calculations without self-energy terms is
 not reasonable because, at least, part of these effects cancels each other.
 From the computational point of view, the higher order retardation terms lead
 to only a small correction to the cumulative BI effect (e.g., see,
 \cite{Indelicato:86}) and, if necessary, they can be taken into account
 posteriori as a first-order perturbation.

\section{Results of atomic DC(B)-calculations.}

 In \Tables{U_CGBG}--\ref{Tl2}, we report the results of calculations for a
 series of heavy elements by the self-consistent field (SCF) method with the
 DC, DCG and DCB Hamiltonians.  The finite-difference (numerical) {\sc HFD}
 code \cite{Bratzev:77} was developed \cite{Tupitsyn:02} to take into account
 the Gaunt and Breit interactions.  The absolute values of one-electron
 energies for most of spinors of the uranium atom are decreased approximately
 on 0.2--0.5\,\% when BI is taken into account in the SCF procedure.  The
 largest change in one-electron energies among the spinors with the same
 principal quantum numbers is observed for the $p_{1/2}$ states (see
 \Table{hfs1}).  The BI contribution to the one-electron energies has the
 opposite sign for the $6d$, $4f$, $5f$ spinors as compared to those for other
 spinors.  The energies of the $5f_{5/2}$ and $5f_{7/2}$ states whose
 occupation numbers are changed in some of transitions between the low-lying
 states of the uranium atom and its compounds are decreased by more than 1\%,
 on 820 and 1000~\cm, correspondingly.  Similar behavior was observed in
 \cite{Lindroth:89c} for the $4f$ and $5d$ spinors of the mercury atom.  The
 inversion of the sign is caused by the change in the electric part (Coulomb
 plus exchange) of the Dirac-Fock potential when spinors are relaxed due to
 the Breit interaction \cite{Lindroth:89c, Kozlov:00}.  The difference between
 matrix elements of the hyperfine interaction in calculations with the DC, DCG
 and DCB Hamiltonians takes place only due to the spinor relaxation.  The
 retardation corrections give about 10\% of the total Breit contributions to
 the one-electron energies.  They can achieve more than 30\% for the hyperfine
 interaction and usually have the opposite sign as compared to the Gaunt
 corrections.

 Table~\ref{U_CGBG} demonstrates that the uranium atom has low-lying states
 with different occupation numbers of the $5f$ shell.  Accounting for the BI
 contribution is especially important for transitions with the excitation of
 $5f$ electrons because of the great contribution of BI to its one-electron
 energies.  The BI contribution to the transition energies is up to 3000~\cm
 (and sometimes is comparable with the DC transition energy).  Similar
 situation is observed for plutonium (\Table{Pu_CGBG}) and is expected for
 other actinides.  Obviously, the cumulative effect caused by BI for other
 calculated atoms grows for elements of the same group with increasing the
 nuclear charge $Z$ and can achieve few hundreds wave numbers for very heavy
 elements (see \Tables{Pb}--\ref{Tl2}).  Nevertheless, the BI between valence
 electrons is decreased with growth of $Z$ mainly because of enlarging the
 average radii of the valence shells as is shown below.

 \section{Disadvantages of DC(B)-based calculations}

 In the previous section, atomic finite-difference SCF calculations based on
 the DC, DCG and DCB Hamiltonians are discussed.  From the formal point of
 view, correlation calculations with these Hamiltonians can provide a very
 high accuracy of physical and chemical properties for molecules containing
 heavy atoms.  However, such calculations are not widely used yet for such
 systems because of the following theoretical and technical complications
 \cite{Visscher:96}:

\begin{itemize}
 \item[{\bf 1}]
      {Too many electrons} are treated explicitly in heavy-atom
      systems and {too large basis set} of Gaussians is required for
      accurate description of the {large number of oscillations},
      which valence spinors have in the presence of heavy atoms.
      Because the number of two-electron integrals grows as the fourth
      degree of the number of basis functions, it leads to serious
      complications of their calculations and, especially, of their
      transformation to a molecular basis.
 \item[{\bf 2}]
      When the Dirac formalism with four-component spinors is used,
      the number of two-electron integrals to be computed is strongly
      increased as compared to the one-component (nonrelativistic or
      scalar-relativistic) and two-component (spin-dependent) cases:
 \begin{itemize}
  \item[(a)]
         The number of kinetically-balanced two-component (``$2c$'')
         uncontracted basis spinors required for description of the
         {\it S}mall components, $N_S^{2c}$, can be estimated as $2N_L^{2c}$,
         where $N_L^{2c}$ is the number of basis spinors for the {\it L}arge
         components; so the total number of the relativistic four-component
         (``$4c$'') basis spinors $N_{bas}^{4c} \sim 3 N_L^{2c}$ and the
         number of calculated two-electron integrals is
         $$
            N_{2eInt}^{4c} \sim (1{+}2{\cdot}2^2{+}2^4) N_{2eInt}^{2c}
                           \equiv 25{\cdot}N_{2eInt}^{2c}\ ,
         $$
         where $N_{2eInt}^{2c}$ is the number of two-electron integrals with
         only the large components when the Coulomb electron--electron
         interaction is taken into account.  Even more computational effort is
         required when the magnetic and retardation electron--electron
         interactions are taken into account. To reduce the number of
         two-electron integrals, different two-component relativistic
         approximations are actively developed first of all to be used in
         molecular calculations (e.g., see \cite{Hess:00}).
  \item[(b)]
         The number of basis $2c$-spinors, $N_{bas}^{2c}$, is twice more than
         the number of nonrelativistic basis {one-component} (``$1c$'')
         orbitals, $N_{bas}^{1c}$, therefore
         $$
            N_{2eInt}^{2c} \sim 2^4/2{\cdot}N_{2eInt}^{1c}
                           \equiv 8{\cdot}N_{2eInt}^{1c}\ ,
         $$
         (the division by factor of two appears due to different symmetry
         properties of two-electron integrals for $1c$ and $2c$ (or $4c$)
         cases)
         and $N_{2eInt}^{2c} \sim 4{\div}6 N_{2eInt}^{2cSO} $, where
         $N_{2eInt}^{2cSO}$ is the number of two-electron integrals in the
         {spin-orbit} basis set which are required to be saved in the computer
         memory.
 \end{itemize}
 \item[{\bf 3}]
      Eq.~(\ref{one}) as well as the Dirac-Fock equations has both negative--
      (positronic) and positive--energy (electronic) solutions. This
      circumstance leads to so called ``continuum dissolution'' when eq.
      (\ref{Heff1}) is applied to electronic structure calculations
      \cite{Brown:51}.  To prevent it, one can use only the positive-energy
      solutions for constructing many-electron basis functions
      (``no--virtual--pair'' approximation).  The way to go beyond the
      no-virtual-pair approximation is to use the normal ordered
      second-quantized representation of eq.~(\ref{Heff1}) with respect to the
      ``vacuum'' level in which all the negative-energy
      states are occupied.  Evaluation of the Dirac-Fock(-Breit)
      bispinors is required for separating the negative--energy and
      positive--energy solutions in the one-particle basis.
      Because the virtual Dirac-Fock(-Breit) bispinors are not optimal for
      precise correlation calculations of molecules, some additional efforts
      can be required to generate an appropriate set of spinors for such calculations.
\end{itemize}

 To avoid complications described in items~$2a$ and~$3$, small components can
 be excluded explicitly from calculations \cite{Hess:00}.  To reduce also the
 difficulties described in items~$1$ and~$2b$, Hamiltonian \eref{Heff1} can be
 replaced by an {effective Hamiltonian} 

\begin{equation}
   {\bf H}^{\rm Ef}\ =\ \sum_{i_v} [{\bf h}^{\rm Schr}(i_v) +
          {\bf U}^{\rm Ef}(i_v)] + \sum_{i_v > j_v} \frac{1}{r_{i_v j_v}}\ .
 \label{Heff2}
\end{equation}
 in practical applications when inner core electrons do not play an active
 role and, besides, the original valence spinors can be smoothed in the
 heavy-atom cores.
 Hamiltonian \eref{Heff2} is written only for valence or ``valence-extended''
 (VE) subspace of electrons denoted by indices $i_v$ and $j_v$ (in the VE
 case, some outermost core shells are also treated explicitly).  ${\bf U}^{\rm
 Ef}$ is an relativistic effective core potential (RECP) or relativistic
 pseudopotential (RPP) operator that is usually written in the
 radially-local (semi-local) \cite{Ermler:88} or separable form (e.g., see
 \cite{Theurich:01} and references).  It simulates, in particular,
 interactions (which can include BI, see below) of the explicitly treated
 VE electrons with those which are excluded from the RECP calculations.
 Besides, the generalized RECP (GRECP) operator \cite{Titov:99,Titov:00} can
 be used, which includes both the radially-local and separable terms
 (additionally, the GRECP method involves possibilities of the Huzinaga-type
 pseudopotential \cite{Bonifacic:74} and other advantages, see below).
 In Eq.~(\ref{Heff1}), ${\bf h}^{\rm Schr}$ is the one-electron Schr\"odinger
 Hamiltonian

\begin{equation}
     {\bf h}^{\rm Schr}\ = - \frac{1}{2} {\vec \nabla}^2 + V^{\rm nuc}\ .
 \label{Schr}
\end{equation}
 Contrary to the four-component wave function used in DC(B) calculations, the
 pseudo-wave function in the (G)RECP case can be both two- and one-component.
 The use of Hamiltonian~(\ref{Heff2}) instead of~(\ref{Heff1}) is, of course,
 an additional approximation and the question about its accuracy appears.  As
 was shown both theoretically and in many calculations, the usual accuracy of
 the radially-local RECP versions is within 1000--3000~\cm for transition
 energies and an appropriate level of accuracy is expected for other chemical
 and physical properties.

\section{ Generalized Relativistic Effective Core Potential.}

 In a series of papers (see \cite{Titov:99, Titov:00} and references), we have
 introduced and developed the GRECP concept which allows one to attain
 practically any desired accuracy, while requiring moderate computational
 efforts.

 The main steps of the scheme of generation of the GRECP version with the
 separable correction are:
\begin{enumerate}
\item   The numerical all-electron relativistic calculation of a generator
        state is carried out for an atom under consideration. For this purpose,
        we use the atomic {\sc HFDB} code~\cite{Bratzev:77,Tupitsyn:02}.
\item   The numerical pseudospinors  {$\widetilde{\varphi}_{nlj}(r)$} are
        constructed of the large components  {$\psi^{l}_{nlj}(r)$} of the
        {outer-core (C)} and valence (V) HFDB spinors so that the innermost
        pseudospinors of them (for each $l$ and $j$) are nodeless, the next
        pseudospinors have one node, and so forth. These pseudospinors satisfy the
        following conditions:

\begin{equation}
 \widetilde{\varphi}_{nlj}(r) =
 \left\{
  \begin{array}{ll}
   \psi^{l}_{nlj}(r),
                                 &  r\geq R_{c}, \\
   y(r)=
   r^{\gamma}\sum_{i=0}^{5}a_{i}r^{i}, &  r<R_{c},
  \end{array}
 \right.
\end{equation}
\[
 \begin{array}{ccc}
  & l=0,1,\ldots,L,~~~~~ j=|l\pm\frac{1}{2}|, & \\
  & n=n_c,n_c',\ldots,n_v, &
 \end{array}
\]
	where  {$L$} is one more than the highest orbital angular 
	momentum of the inner core (IC) spinors. The leading 
	power  {$\gamma$} in the polynomial is typically chosen 
	to be close to {$L+1$} in order to ensure a sufficient 
	ejection of the valence and outer-core electrons from the IC 
	region.	 The $a_i$ coefficients are determined by 
	the following requirements: 
	\begin{itemize}
	\item    {$\{\widetilde{\varphi}_{nlj}\}$} set is orthonormalized,
        \item    {$y$} and its first four derivatives match
		 {$\psi^{l}_{nlj}$} and its derivatives,
        \item    {$y$} is a smooth and nodeless function, and
        \item    {$\widetilde{\varphi}_{nlj}$} ensures a sufficiently
                smooth shape of the corresponding potential.
	\end{itemize}
         {$R_{c}$} is chosen near the extremum of the spinor so 
	that the corresponding pseudospinor has the defined above number 
	of nodes. In practice, the  {$R_{c}$} radii for the 
	different spinors should be chosen close to each other  
	to generate smooth potentials.
\item   The  {$U_{nlj}$} potentials are derived for each 
        {$l{=}0,\ldots,L$} and  {$j{=}|l \pm \frac{1}{2}|$}        
        for the valence and outer-core pseudospinors so that
	the {$\widetilde{\varphi}_{nlj}$} are solutions of
	the nonrelativistic-type Hartree-Fock equations in the
	{\it jj}-coupling scheme for a ``pseudoatom'' with
        the removed {IC} electrons.
\begin{eqnarray}
  U_{nlj}(r)  =  \widetilde{\varphi}_{nlj}^{-1}(r)
                \Biggl[\Biggl( \frac{1}{2} 
		{\bf \frac{d^{2}}{dr^{2}} }
                - \frac{l(l+1)}{2r^{2}}
                + \frac{Z_{ic}}{r} 
		 -   \widetilde{\bf J}(r) 
	          +  \widetilde{\bf K}(r)
       	  +  \varepsilon_{nlj}  \Biggr) \widetilde{\varphi}_{nlj}(r)  
	  + 
                \sum_{n'\neq n} \widetilde{\varepsilon}_{n'nlj}
                \widetilde{\varphi}_{n'lj}(r) \Biggr] 
          ,
\label{U_nlj}
\end{eqnarray}
 where {$Z_{ic}$} is the charge of the nucleus decreased by the number of
 {IC} electrons, {$\widetilde{\bf J}$} and {$\widetilde{\bf K}$} are Coulomb
 and exchange operators calculated with the {$\widetilde{\varphi}_{nlj}$}
 pseudospinors, {$\varepsilon_{nlj}$} are the one-electron energies of the
 corresponding spinors, and  {$\widetilde{\varepsilon}_{n'nlj}$} are
 off-diagonal Lagrange multipliers (which are, in general, slightly different
 for the original bispinors and pseudospinors).

 \item The GRECP operator with the separable correction written in the spinor
       representation~\cite{Titov:99} is as

\begin{eqnarray}
 \label{UGRECP}
 {\bf U}  &=&  U_{n_vLJ}(r) 
                 +  \sum_{l=0}^L \sum_{j=|l-1/2|}^{l+1/2}
		   \Bigl\{\bigl[U_{n_vlj}(r) 
		    -  U_{n_vLJ}(r)\bigr]
                   {\bf P}_{lj}   \nonumber\\
                &+&   \sum_{n_c} 
                   \bigl[U_{n_clj}(r) 
		   -  U_{n_vlj}(r)\bigr] 
                   \widetilde{\bf P}_{n_clj} 
		   +   \sum_{n_c}  \widetilde{\bf P}_{n_clj}
                   \bigl[U_{n_clj}(r) 
		    -  U_{n_vlj}(r)\bigr] \\ 
              &-&   \sum_{n_c,n_c'} 
                   \widetilde{\bf P}_{n_clj}
                   \biggl[\frac{U_{n_clj}(r)+U_{n_c'lj}(r)}{2} 
		      -   U_{n_vlj}(r)\biggr]
                   \widetilde{\bf P}_{n_c'lj}\Bigr\}, \nonumber
\end{eqnarray}
 where
\[
  {\bf P}_{lj} = \sum_{m=-j}^j
    \bigl| ljm \bigl\rangle \bigr\langle ljm \bigr|,
\ \ \ \ \ \ \ \ \ \
  \widetilde{\bf P}_{n_clj} = \sum_{m=-j}^j
  \bigl| \widetilde{n_cljm} \bigl\rangle \bigr\langle \widetilde{n_cljm} \bigr|,
\]
         {$\bigl| ljm \bigl\rangle \bigr\langle ljm \bigr|$} 
 is the projector on the two-component spin-angular function 
	 {$\chi_{ljm}$}, 
         {$\bigl| \widetilde{n_cljm} \bigl\rangle
                \bigr\langle \widetilde{n_cljm} \bigr|$}
 is the projector on the outer core pseudospinors 
         {$\widetilde{\varphi}_{n_clj}\chi_{ljm}$}, 
	and  {$J=L+1/2$}.
\end{enumerate}
 Two of the major features of the GRECP version with the separable correction
 described here are generating of the effective potential components for
 pseudospinors which may have nodes, and addition of non-local separable terms
 with projectors on the outer core pseudospinors (the second and third lines
 in \Eq{UGRECP}) to the standard semi-local RECP operator (the first line in
 \Eq{UGRECP}).  Some other GRECP versions are described and discussed in
 papers \cite{Titov:99,Titov:00,Titov:02Dis} in details.

 The GRECP operator in spinor representation \eref{UGRECP} is mainly used in
 our atomic calculations.  The spin-orbit representation of this operator
 which can be found in~\cite{Titov:99} is more efficient
 in practice being applied to
 molecular calculations.  Despite the complexity of expression \eref{UGRECP}
 for the GRECP operator, the calculation of its one-electron integrals
 is not significantly more expensive than that for the case of the
 standard radially-local RECP operator.

 \section{ Accounting for Breit effects in GRECP}

 Let us consider the contributions of BI between electrons from different
 shells to the total energy of a heavy atom~\cite{Titov:02Dis}.  The following
 estimate can be applied (e.g., see \cite{Labzowsky:93c})

\[
  \langle a,b | ({\vec \alpha}_i{\cdot}{\vec \alpha}_j) | a,b \rangle \sim
      \frac{1}{c^2} \langle ({\vec v}_a{\cdot}{\vec v}_b) \rangle\ .
\]
 For an uncoupled one-electron state $a$ one has
\[
  \langle a|\vec{\alpha}|a \rangle \sim \frac{\langle\vec{v}\rangle_a}{c}\ ,
  \qquad  \frac{|\langle\vec{v}\rangle_a|}{c} \sim \alpha Z_a^*\ ,
\]
 where $Z_a^*$ is an effective charge of the core that is experienced by the electron
 in the $a$-th state, $Z_a^*{=}Z{-}N_c^a$, $Z$ is the nuclear charge, $N_c^a$
 is the number of core electrons with respect to the $a$-th state (the core
 radius $R_c^a \sim \langle a|r|a \rangle$).  Besides,
 $\langle1{/}r_{12}\rangle$ can be estimated as $\langle 1{/}r \rangle$ for
 the outermost (from $a$ and $b$) one-electron state:
$$ \langle ab| \frac{1}{r_{12}} |ab \rangle \sim \min\left[ \langle
a|\frac{1}{r}|a \rangle, \langle b|\frac{1}{r}|b \rangle \right] \sim
\min\left[ Z_a^*, Z_b^* \right]\ .  $$ Thus, the absolute value of BI between
electrons in states $a$ and $b$ can be roughly estimated as

\begin{equation}
  \left|B_{ab}\right|  \sim  \alpha^2 Z_a^* Z_b^* \cdot
         \min\left[ Z_a^*,Z_b^* \right]\ .
\label{Bsim}
\end{equation}
 When \Eq{Heff2} is applied instead of \Eq{Heff1}, it means neglecting (or, in
 some sense, freezing) the BI between the VE electrons.  In the case of
 $a{\equiv}v$ and $b{\equiv}v'$, one has $Z_{a,b}^*{\sim}1$ and
\[
   \left|B_{vv'}\right| \sim \alpha^2\
            \approx \frac{1}{2}{\cdot}10^{-4}\ {\rm a.u.}
            \approx 10\ {\rm cm}^{-1}\ .
\]
 Thus, this contribution is negligible for the ``chemical accuracy'' (about
 350~\cm) of calculations.  This is demonstrated in \Tables{contr1}
 and~\ref{contr2}, where we calculated the BI contributions to the total and
 transition energies of the uranium atom.  In these calculations we have
 subdivided all the electrons on the ``core'' and ``valence'' subspaces by
 different ways and the BI contributions between the valence electrons are
 neglected for each of such subdivisions (i.e., only core--core and
 core--valence contributions are taken into account).  We have also calculated
 the BI contributions to the total energy between the valence--valence,
 core--valence and core--core shells and have estimated them for different $s$
 shells applying \Eq{Bsim}.  The results are collected in \Table{valence}.  In
 this table we have calculated $Z_a^*$ as $\langle a|1{/}r|a \rangle$ and have
 introduced the normalizing factors $C_{N}=0.34$ as some average multiplier
 for the BI contributions (``direct'' plus ``exchange'') per an electron pair
 within the considered $s$ shells (see the footnote to \Table{valence} for
 more details). It may be seen from \Table{valence} that the estimates from
 \eref{Bsim} agree with the calculated values mainly within the factor of two.
 The suggested estimate is expected to be substantially cruder for higher
 total electronic momenta.

\begin{figure}
\includegraphics[scale=1]{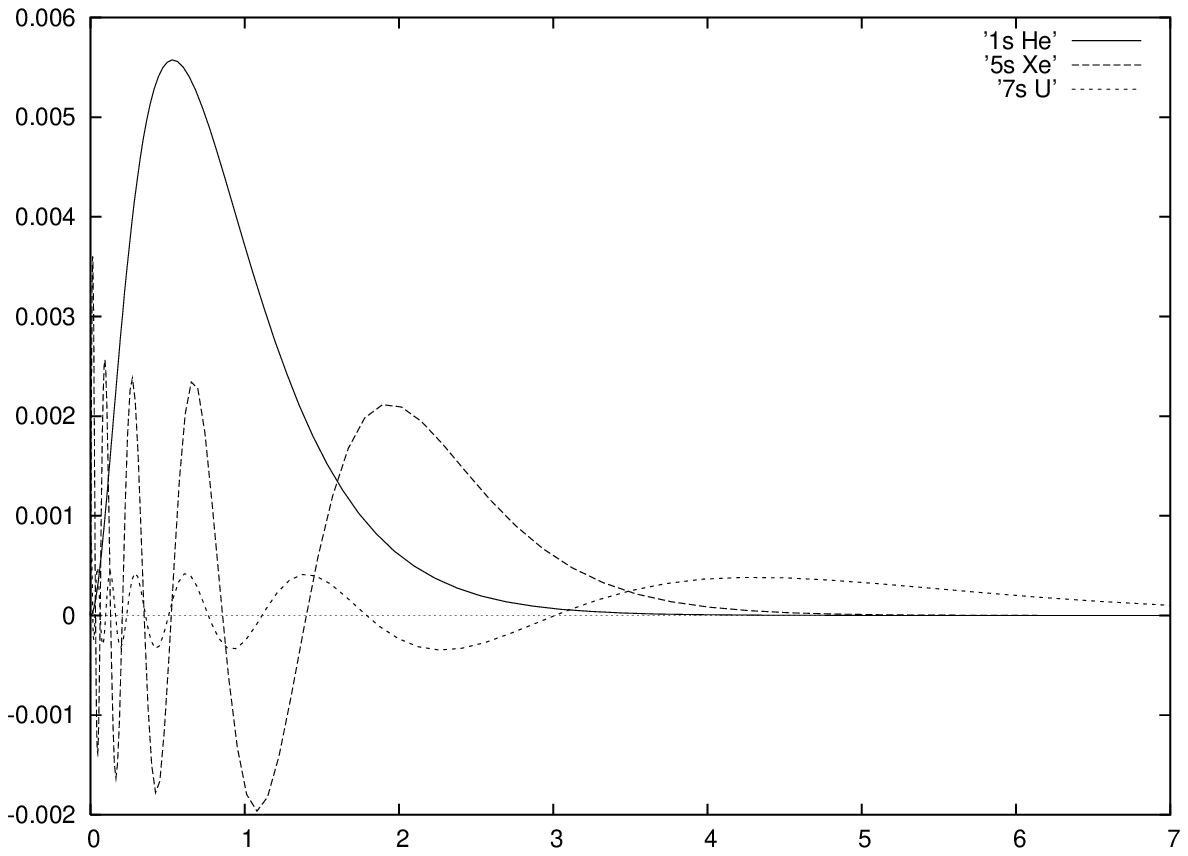}
\includegraphics[scale=1]{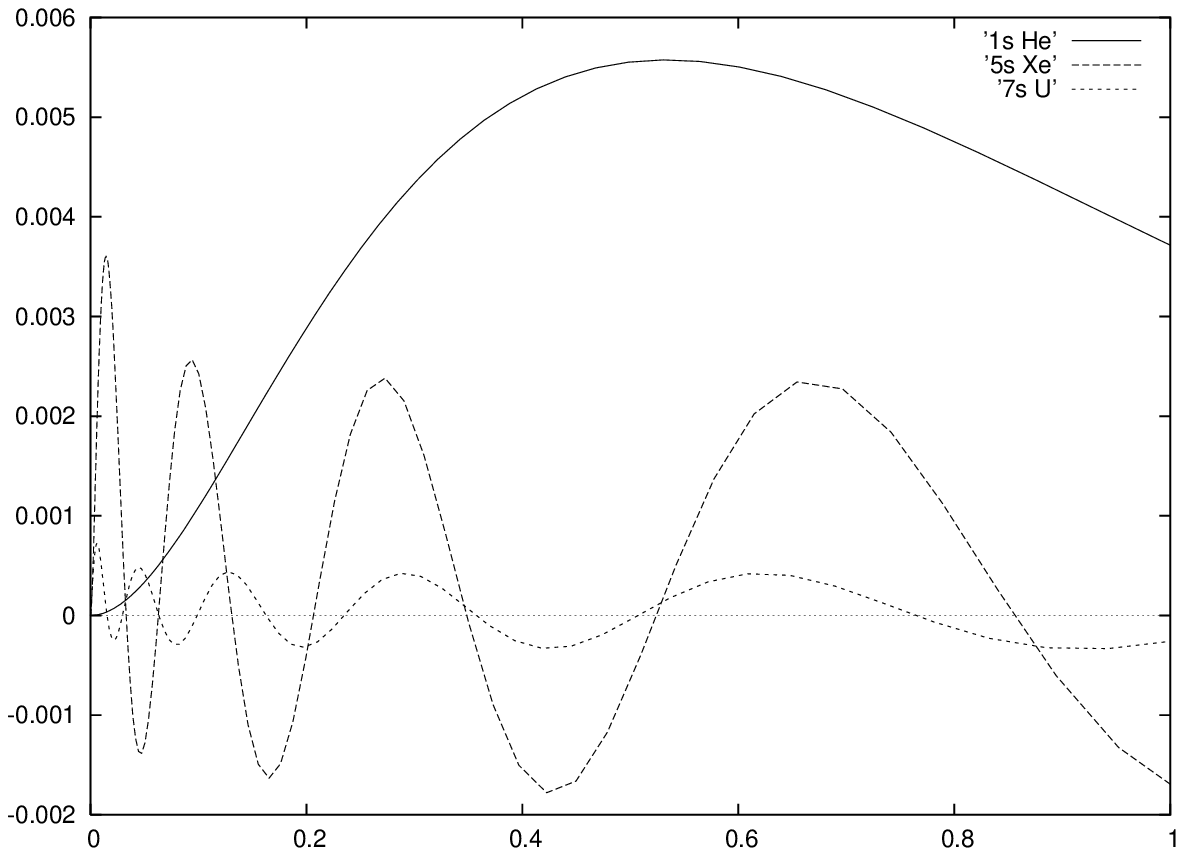}
\caption{\label{osc} Product
 $r\psi^{l}_{ns}(r)\cdot r\psi^{s}_{ns}(r)$
 as a function of $r$ for valence $s$ functions of different atoms (in a.u.).}
\end{figure}

 The valence--valence contribution is usually decreased with the increase of
 $Z$.  There are several reasons for such behaviour.
 The expression for BI contains double integration on radial variables of
 the products of large and small radial components of Dirac bispinor

\begin{equation}
   \int \limits_{0}^{\infty} dr_1  \int \limits_{0}^{\infty} dr_2 \,
   \psi^{l}_{a}(r_1)  \psi^{l}_{c}(r_2) \, u_{k}(r_1,r_2) \,
   \psi^{s}_{b}(r_1)  \psi^{s}_{d}(r_2) r_1^2r_2^2\ ,
 \label{rad1}
\end{equation}
%
 where $u_{k}(r_1,r_2) = r_<^{k}/r_>^{k+1}$, 
 $r_<=\min[r_1,r_2]$ and $r_>=\max[r_1,r_2]$.  
 For high $Z$ radial parts of the large and small components of the valence
 spinors have a number of oscillations, which usually do not coincide.  It
 leads to oscillations in the products
 $\psi^{l}_{a}(r)\cdot\psi^{s}_{c}(r)\cdot r^2$ (see Fig.~\ref{osc}) and
 large cancellation in integral~(\ref{rad1}) as compared to the case of
 integrating over the amplitude of the integrant (see \Table{valence}): 

\begin{equation}
   \int \limits_{0}^{\infty} dr_1  \int \limits_{0}^{\infty} dr_2 \,
   \left|\, \psi^{l}_{a}(r_1)  \psi^{l}_{c}(r_2) \, u_{k}(r_1,r_2) \,
   \psi^{s}_{b}(r_1)  \psi^{s}_{d}(r_2)\,\right|\, r_1^2r_2^2\ .
 \label{rad2}
\end{equation}
 In particular, it also means that the ``incompensated'' BI contribution
 between the valence electrons is mainly taken in the valence region.  As one
 can see from \Fig{osc}, the absolute value of the
 $r\psi^{l}_{a}(r)\cdot r\psi^{s}_{c}(r)$
 is decreased with the increase
 of $Z$.  It may be easily understood when the kinetic balance condition is
 applied to derive the small components from the large ones in the valence
 region:

\begin{equation}
   r\cdot\psi^{s}(r)\  =\  \frac{\alpha}{2}
    \left(\frac{d}{dr}+\frac{k}{r}\right)\left[r\cdot\psi^{l}(r)\right]\ .
 \label{s_l}
\end{equation}
 The atomic radius is increased with the increase of $Z$ and, thus, the large
 components of the valence spinors become smaller and smoother.  It leads to
 the decrease of the small components in accord to eq.~(\ref{s_l}).

 The core electrons can be considered as frozen when studying majority of
 physical-chemical properties and processes of practical interest. Since they
 usually form closed shells in calculations, the total contribution of the
 direct part of BI between the core and valence electrons is equal to zero and
 only the exchange part of this interaction gives nonzero contribution.  For
 the innermost $1s$-shell ($a{\equiv}c$) of a heavy atom, one has
 $Z_a^*{\sim}Z{\sim}100$ by the order of magnitude and

\[
    \left|B_{cv}\right| \sim \alpha^2{\cdot}100
            \approx \frac{1}{2}{\cdot}10^{-2}\ {\rm a.u.}
            \approx 1000\ {\rm cm}^{-1}\ ,
\]
 that is quite essential for calculations on the level of ``chemical
 accuracy''.  (We found in the present calculations that contribution from the
 exchange interaction is usually larger than the direct one in the cases of
 pairs of spinors
 $(ns_{1/2},ms_{1/2})$; $(ns_{1/2},mp_{1/2})$; $(ns_{1/2},mp_{3/2})$;
  $(np_{1/2},mp_{1/2})$; $(np_{1/2},md_{3/2})$
 and smaller in the cases of pairs of spinors
 $(ns_{1/2},md_{3/2})$; $(ns_{1/2},md_{5/2})$; $(ns_{1/2},mf_{5/2})$;
  $(ns_{1/2},mf_{7/2})$; $(np_{1/2},mp_{3/2})$; $(np_{1/2},md_{5/2})$;
  $(np_{1/2},mf_{7/2})$; $(np_{3/2},md_{3/2})$; $(np_{3/2},mf_{5/2})$;
  $(np_{3/2},mf_{7/2})$.)
 Therefore, one can just take into account the $B_{cv}$ contributions (and
 neglect the $B_{vv'}$ terms) when generating (G)RECPs with only valence
 electrons treated explicitly, where index $c$ runs over the core electrons
 only. The latter are explicitly excluded from the (G)RECP calculations,
 therefore, the effective operator for $B_{cv}$ acting on the valence shells
 has the same spin-angular dependence as the conventional radially-local RECP
 has.  (For higher accuracy, GRECP can be used, which, besides, can
 efficiently treat other QED effects even for highly charged ions of
 heavy-element systems.) Thus, BI can be taken into account directly when the
 {\sc HFDB} calculations are performed (see item~1 in the previous section) to
 generate valence bispinors and their one-electron energies, but in the
 procedure of inversion of the two-component HF equations for generating the
 (G)RECP components (see item~3 in the previous section), the conventional
 interelectronic Coulomb interaction can be used instead of the Coulomb-Breit
 one.  Afterwards, one should consider only the Coulomb interaction between
 the explicitly treated electrons in the correlation (G)RECP calculations.  In
 principle, the errors due to neglecting BI between the VE electrons (as well
 as the errors in reproducing the interelectronic Coulomb interaction due to
 spinor smoothing) can be compensated with good accuracy by the term-splitting
 GRECP correction \cite{Titov:99, Titov:00}.  Alternatively, the procedure of
 restoration of proper shapes for the four-component VE spinors in the
 heavy-atom cores \cite{Titov:96, Petrov:02} after the molecular GRECP
 calculation can be applied, then both BI between the VE electrons and the
 proper interelectronic Coulomb interaction are directly taken into account
 with the restored wave function.

 The Breit interaction for both the core electrons, $B_{cc'}$, can be of the
 same order of magnitude as the Coulomb interaction between them.  However,
 $B_{cc'}$ does not contribute to ``differential'' (valence) properties
 directly, whereas its indirect contribution can be easily taken into account
 with the help of the (G)RECP technique.

\section{Results of the GRECP calculations}

 In the given section we consider accounting for BI by the GRECP method only
 for actinides for which the Breit effects can attain a few thousands wave
 numbers even to the energies of lowest-lying electronic excitations.
 Accounting for BI in different RECP and GRECP versions for superheavy
 elements and actinides will be considered in details in our forthcoming paper
 \cite{Mosyagin:03}.  For the uranium and plutonium atoms we constructed
 different 24-- and 26--electrons GRECPs, correspondingly, which effectively
 account for the Breit effects.  The GRECP($f^3$) and GRECP($f^2$) for the
 uranium atom were constructed using the generator states with the occupation
 numbers of the $5f$-shell equal to~3 and~2.  In the case of
 the plutonium atom the configurations
 with the occupation numbers of the $5f$-shell equal to~6 and~5 were used,
 respectively.  The results of the GRECP/SCF calculations as compared to those
 of all-electron HFDB calculations are presented in \Tables{U_CGBG}
 and~\ref{Pu_CGBG}.

 The errors of calculations with these GRECPs can be collected into two
 groups.  The errors for transitions without the change in the occupation
 number of the $5f$ shell are rather small.  The errors for transitions with
 the change in the occupation number of this shell can achieve few thousand
 wave numbers.  The latter errors have a systematic nature and are mainly
 connected with two facts: {\it (a)} the $5f$ shell of uranium and plutonium
 is described with the help of nodeless pseudospinors in the present GRECP
 versions and {\it (b)} the relaxation of the [Xe\,$4f^{14}$]-like core shells
 is not taken into account since they are excluded from the GRECP calculations
 explicitly.  These errors can be reduced significantly if one includes the
 $4f^{14} 5s^2 5p^6$ electrons in the calculations for uranium and plutonium
 explicitly.  Another way is to apply the self-consistent (SfC) GRECP version
 \cite{Titov:95,Titov:99}.  In the SfC~GRECP case, the circumstance is taken
 into account that the GRECPs generated for different occupation numbers of
 the outermost core shell(s) ($N_{5f}^{\rm occ}$ in our case) are somewhat
 distinguished due to above mentioned reasons (see \cite{Titov:99,Titov:02Dis}
 for more details).  The dependence on $N_{5f}^{\rm occ}$ is introduced
 explicitly to the SfC~GRECP operators, which are constructed for uranium and
 plutonium on the basis of the above generated GRECPs. 
 As it can be seen from {\Tables{U_CGBG}} and~\ref{Pu_CGBG}, the SfC~GRECPs
 allow one to increase the accuracy of the calculations without extension of
 the space of explicitly treated electrons up to an order of magnitude and
 more.

 \vspace{3mm}
\acknowledgments
 The present work is supported by the U.S.\ CRDF Grant No.\ RP2--2339--GA--02
 and by the RFBR grant 03--03--32335.  A.P.\ is grateful to Ministry of
 education of Russian Federation (grant PD\,02--1.3--236) and to St-Petersburg
 Committee of Science (grant PD\,03-1.3-60).
 N.M.\ is supported by the scientific fellowship grant of the governor of
 Leningrad district.


\begin{table}
\begin{center}
\caption{ 
         Energies of transitions between the states averaged over
         nonrelativistic configurations of the uranium atom 
         calculated by the DHF method with the Coulomb, Coulomb-Gaunt and
         Coulomb-Breit two-electrons interactions taken into account and by
         the SCF method with different GRECPs (in \cm). }
\vspace{0.1cm}
\begin{tabular}{lrrr|rrr}
\hline
\hline
       &             & Gaunt       & Retard.\    & \multicolumn{3}{c}{Absolute error of 24e-GRECPs }  \\
       &             & contr.\     & contr.\     &  (GRECP($f^3$) & (GRECP($f^2$) & (SfC GRECP    \\
 Transition  & DC          & (DCG-DC)   & (DCB-DCG)   & -DCB)        & -DCB)          &    -DCB)   \\
\hline
$5f^3 7s^2 6d^1 \rightarrow $ &          &         &       &       &      &     \\
$5f^3 7s^2 7p^1             $ &    7424  &  105    &  -13  & -37   & -191 & -41 \\
$5f^3 7s^2                  $ &   36227  &  67     &   -5  & 1     & -109 & -3  \\
$5f^3 7s^1 6d^2             $ &   13202  &  -84    &    6  & -5    & 57   & -4  \\
$5f^3 7s^1 6d^1 7p^1        $ &   17214  &  -11    &   -3  & -29   & -81  & -29 \\
$5f^3 7s^1 6d^1             $ &   42373  &  -47    &    2  & -15   & -29  & -15 \\
$5f^3 6d^2                  $ &   54714  &  -147   &    9  & -11   & 54   & -9  \\
$5f^3 7s^2 6d^1 \rightarrow $ &          &         &       &       &      &     \\
$5f^4 7s^2                  $ &   16408  &  -736   &  110  & -424  & 125  & -10 \\
$5f^4 7s^2      \rightarrow $ &          &         &       &       &      &     \\
$5f^4 7s^1 6d^1             $ & 15053    &  -44    &    1  & 34    & 63   & -3  \\
$5f^4 7s^1 7p^1             $ & 14953    &  -19    &   -2  & -68   & -112 & -25 \\
$5f^4 7s^1                  $ & 38863    &  -54    &    4  & -33   & -43  & -21 \\
$5f^4 6d^2                  $ & 33874    &  -84    &    2  & 63    & 109  & -1  \\
$5f^4 6d^1 7p^1             $ & 32194    &  -80    &    1  & 3     & 11   & -11 \\
$5f^4 6d^1                  $ & 53488    &  -114   &    6  &  26   & 65   & -15 \\
$5f^3 7s^2 6d^1 \rightarrow $ &          &         &       &       &      &     \\
$5f^2 7s^2 6d^2             $ &    3859  &  914    & -135  & 986   & 311  & -20 \\
$5f^2 7s^2 6d^2 \rightarrow $ &          &         &       &       &      &     \\
$5f^2 7s^2 6d^1 7p^1        $ &   12690  & 135     & -17   & 164   & -13  & -5  \\
$5f^2 7s^2 6d^1             $ &  42710   & 91      & -8    & 150   & 23   & 32  \\
$5f^2 7s^1 6d^3             $ &  10584   & -113    &  9    & -103  & -20  & -24 \\
$5f^2 7s^1 6d^2 7p^1        $ &  19232   & -11     & -4    & 28    & -28  & -27 \\
$5f^2 7s^1 6d^2             $ &  45402   & -52     &  2    & -4    & -17  & -15 \\
$5f^2 6d^3                  $ &  54780   & -180    & 11    & -115  & -27  & -32 \\
$5f^3 7s^2 6d^1 \rightarrow $ &          &         &       &       &      &     \\
$5f^1 7s^2 6d^3             $ & 29773    & 1964    & -290  & 2728  & 1285 & -141\\
$5f^1 7s^2 6d^3 \rightarrow $ &          &         &       &       &      &     \\
$5f^1 7s^2 6d^2 7p^1        $ & 18189    & 156     & -19   & 432   & 242  & 82  \\
$5f^1 7s^2 6d^2             $ & 49233    & 106     & -10   & 340   & 204  & 99  \\
$5f^1 7s^1 6d^4             $ & 7455     & -134    &  11   & -256  & -157 & -80 \\
$5f^1 7s^1 6d^3 7p^1        $ & 21056    & -13     &  -5   & 93    & 37   & -17 \\
$5f^1 7s^1 6d^3             $ & 48058    & -59     &   2   & -7    & -15  & -24 \\
$5f^1 6d^4                  $ & 54003    & -210    &  14   &  -284 & -181 & -100\\
$5f^3 7s^2 6d^1 \rightarrow $ &          &         &       &       &      &     \\
$5f^5                       $ & 100587   & -1313   & 187   &  -534 & 392  & 14  \\
\hline
\end{tabular}
\label{U_CGBG}
\end{center}
\end{table}

\begin{table}
\begin{center}
\caption{ 
         Energies of transitions between the states averaged over
         nonrelativistic configurations of the plutonium atom 
         calculated by the DHF method with the Coulomb, Coulomb-Gaunt and
         Coulomb-Breit two-electrons interactions taken into account and by
         the SCF method with different GRECPs (in \cm). }
\vspace{0.1cm}
\begin{tabular}{lrrr|rrr}
\hline
\hline
       &             & Gaunt       & Retard.\    & \multicolumn{3}{c}{Absolute error of 26e-GRECPs }  \\
       &             & contr.\     & contr.\     &  (GRECP($f^6$) & (GRECP($f^5$) & (SfC GRECP      \\
 Transition  & DC          & (DCG-DC)   & (DCB-DCG)   & -DCB)        & -DCB)          &    -DCB)  \\
\hline
$ 5f^6 7s^2 \rightarrow      $  &       &       &        &      &       &        \\
$ 5f^6 7s^1 6d^1             $  & 17217 & -56   &  3     &  12  & 32    & 10     \\    
$ 5f^6 7s^1 7p^1             $  & 15697 & -16   &  -3    & -27  & -57   & -26    \\   
$ 5f^6 7s^1                  $  & 39900 & -50   &  3     & -16  & -27   & -15    \\   
$ 5f^6 6d^1                  $  & 56908 & -120  &  6     & 8    & 31    &  8     \\   
$ 5f^6 7p^1                  $  & 66747 & -70   &  0     & -32  & -86   & -31    \\
$ 5f^6 7s^2 \rightarrow      $  &       &       &        &      &       &        \\                            $ 5f^7 7s^1                  $  & 44194 & -587  &  84    & -450 & -119  & -70    \\   
$ 5f^7 7s^1 \rightarrow      $  &       &       &        &      &       &        \\              
$ 5f^7 6d^1                  $  & 19931 & -58   &   4    & 0   & 9     & -4      \\ 
$ 5f^7 7p^1                  $  & 14878 & -68   &  6     & -29  & -26   & -17    \\
$ 5f^7                       $  & 35053 & -106  & 10     & -23  &  0    & -24    \\
$ 5f^6 7s^2 \rightarrow      $  &       &       &        &      &       &        \\                
$ 5f^5 7s^2 6d^1             $  & -3803 & 826   &  -122  & 960  & 497   & 6      \\ 
$ 5f^5 7s^2 6d^1 \rightarrow $  &       &       &        &      &       &        \\
$ 5f^5 7s^2 7p^1             $  & 6650  & 105   &  -12   & 72   & -26   & -25    \\
$ 5f^5 7s^1 6d^2             $  & 15125 & -87   & 6      & -42  & -3    & -4     \\ 
$ 5f^5 7s^1 6d^1 7p^1        $  & 18262 & -13   & -3     & 19   & -14   &  -14   \\
$ 5f^5 7s^2                  $  & 35849 & 65    & -4     & 74   & 3     & 4      \\
$ 5f^5 7s^1 6d^1             $  & 43812 & -51   & 3      & 3    & -7    & -6     \\
$ 5f^6 7s^2 \rightarrow      $  &       &       &        &      &       &        \\         
$ 5f^4 7s^2 6d^2             $  & 15880 & 1812  & -267   & 2585 & 1580  & -124   \\
$ 5f^4 7s^2 6d^2 \rightarrow $  &       &       &        &      &       &        \\
$ 5f^4 7s^2 6d^1 7p^1        $  & 12319 & 130   & -15    & 273  & 164   & 63     \\
$ 5f^4 7s^1 6d^3             $  & 12325 & -112  & 8      & -146 & -94   & -49    \\
$ 5f^4 7s^1 6d^2 7p^1        $  & 20424 & -14   & -5     & 75   & 42    & 8      \\
$ 5f^4 7s^2 6d^1             $  & 42764 & 84    & -7     & 219  & 140   & 71     \\
$ 5f^4 7s^1 6d^2             $  & 47004 & -57   & 2      & 11   & 3     & -5     \\
$ 5f^6 7s^2 \rightarrow      $  &       &       &        &      &       &        \\  
$ 5f^3 7s^2 6d^3             $  & 60152 & 2927  & -431   & 5022 & 3412  & -411   \\
$ 5f^3 7s^2 6d^3 \rightarrow $  &       &       &        &      &       &        \\
$ 5f^3 7s^1 6d^4             $  & 9050  & -134  & 10     & -295 & -235  & -136   \\ 
$ 5f^3 7s^2 6d^2 7p^1        $  & 18114 & 151   & -18    & 533  & 416  & 208    \\
$ 5f^3 7s^1 6d^3 7p^1        $  & 22344 & -16   & -5     & 140  & 107  & 34     \\
$ 5f^3 6d^5                  $  & 24372 & -251  & 19     & -522 & -413  & -230   \\
$ 5f^3 7s^2 6d^2             $  & 49588 & 97    & -8     & 399  & 314  & 174    \\
$ 5f^3 7s^1 6d^3             $  & 49757 & -65   & 2      & 7    & 1  & -15    \\
\hline
\end{tabular}
\label{Pu_CGBG}
\end{center}
\end{table}

\begin{table}
\begin{center}
\caption{ Diagonal matrix elements of the hyperfine interaction (in MHz) and
          one-electron energies (in a.u.)
          obtained by the Dirac-Hartree-Fock method with the Coulomb,
          Coulomb-Gaunt and Coulomb-Breit two-electron interactions taken
          into account in calculation of the $5f^3 7s^2 6d^1$ configuration of
          the $^{235}$U atom.}
\vspace{0.3cm}
\begin{tabular}{crrr|ddd}
\hline
\hline
       & \multicolumn{3}{c}{Hyperfine interaction} & \multicolumn{3}{c}{One-electron energy} \\
spinor & \multicolumn{1}{c}{DC} & \multicolumn{1}{c}{DCG} & \multicolumn{1}{c}{DCB} & \multicolumn{1}{c}{DC} & \multicolumn{1}{c}{DCG} & \multicolumn{1}{c}{DCB}  \\
       & \multicolumn{1}{c}{} & \multicolumn{1}{c}{-DC} & \multicolumn{1}{c}{-DCG} & \multicolumn{1}{c}{} & \multicolumn{1}{c}{-DC} & \multicolumn{1}{c}{-DCG}  \\
\hline
 $1s$        &~~ -36798227 &~~~~ 133035 & ~~~~~~ 7027 & -4279.2641972 & 19.1907227 & -1.3823594  \\
 $2s$        &    -5773435 &      14022 &      -4873  &  -806.1661987 &  2.3414060 &  -.2158518  \\
 $3s$        &    -1355941 &      4565  &      -1898  &  -206.6114745 &   .4637954 &  -.0453239  \\
 $4s$        &    -364146  &      1568  &      -613   &   -54.3244989 &   .1054092 &  -.0110359  \\
 $5s$        &    -89193.1 &      406.9 &      -151.3 &   -12.5963766 &   .0217699 &  -.0021773  \\
 $6s$        &    -17527.1 &       81.0 &      -28.8  &    -2.1379683 &   .0038303 &  -.0003502  \\
 $7s$        &    -1595.43 &     5.72   &      -2.16  &     -.2023518 &   .0002373 &  -.0000114  \\
             &             &            &             &               &            &             \\
 $2p_{1/2} $ &    -1967173 &     20229  &      -1243  &  -776.3775115 &  3.9620074 &  -.3915975  \\
 $3p_{1/2} $ &    -473306  &     4570   &       -516  &  -193.1006247 &   .7876421 &  -.0776144  \\
 $4p_{1/2} $ &    -124395  &     1263   &      -171   &   -48.1943130 &   .1833323 &  -.0187803  \\
 $5p_{1/2} $ &    -28600.0 &     296.6  &      -39.4  &   -10.1279114 &   .0375639 &  -.0037016  \\
 $6p_{1/2} $ &    -4887.86 &     55.87  &      -7.01  &    -1.3430981 &   .0059178 &  -.0005425  \\
 $2p_{3/2} $ &    -136925  &     1266   &      -246   &  -635.5705743 &  2.5069539 &  -.3943068  \\
 $3p_{3/2} $ &    -33035.1 &      275.9 &      -69.4  &  -160.3225666 &   .4802565 &  -.0777980  \\
 $4p_{3/2} $ &    -8678.45 &      74.70 &      -20.67 &   -39.5412400 &   .0999325 &  -.0177798  \\
 $5p_{3/2} $ &    -1958.34 &     16.87  &      -4.75  &    -8.0927441 &   .0177925 &  -.0035444  \\
 $6p_{3/2} $ &    -300.545 &     2.773  &      -.767  &     -.9846770 &   .0022487 &  -.0005017  \\
             &             &            &             &               &            &             \\ 
 $3d_{3/2}$  &    -17399.0 &     94.4   &      -11.9  &  -139.0184571 &   .3490051 &  -.0371458  \\
 $4d_{3/2}$  &    -4214.13 &     22.32  &      -3.95  &   -29.7336816 &   .0559536 &  -.0058241  \\
 $5d_{3/2}$  &    -780.353 &     3.686  &      -.691  &    -4.3513548 &   .0051781 &  -.0004503  \\
 $6d_{3/2}$  &    -49.0723 &     .1622  &      -.0351 &     -.1927517 &  -.0002264 &   .0000227  \\
 $3d_{5/2}$  &    -6517.26 &     29.25  &      -5.78  &  -132.4187434 &   .2429070 &  -.0357106  \\
 $4d_{5/2}$  &    -1568.65 &     6.29   &      -1.75  &   -28.1305532 &   .0306448 &  -.0055668  \\
 $5d_{5/2}$  &    -286.521 &     .910   &      -.306  &    -4.0416478 &   .0004606 &  -.0004474  \\
 $6d_{5/2}$  &    -16.9990 &     .0176  &      -.0158 &     -.1832747 &  -.0003916 &   .0000202  \\
             &             &            &             &               &            &             \\
 $4f_{5/2}$  &    -773.708 &      1.446 &      -.190  &   -15.2045521 &  -.0051455 &   .0026720  \\
 $5f_{5/2}$  &    -62.1555 &     -.1984 &      .0271  &     -.3470475 &  -.0044479 &   .0007116  \\
 $4f_{7/2}$  &    -410.285 &     .465   &      -.143  &   -14.7923447 &  -.0166711 &   .0026479  \\
 $5f_{7/2}$  &    -32.0892 &     -.1674 &      .0122  &     -.3195979 &  -.0052414 &   .0006863  \\
\hline
\hline
\end{tabular}
\label{hfs1}
\end{center}
\end{table}

\begin{table}
\caption{Transition energies between states with $ns^2np^2$ configuration
 of the Tin ($Z=50$), Lead ($Z=82$) and Ekalead ($Z=114$) atoms calculated 
 by Dirac-Hartree-Fock method with Coulomb and Coulomb-Gaunt two-electron 
 interactions. All values (excepting the relative differences) are in \cm.}

\begin{tabular} {lrrrr}
\hline
\hline
\multicolumn{5}{c}{  Tin atom} \\
\hline
\hline
  configuration, &    DC &   DCG &   absolute &    relative \\ 
  term
  &  &  &   difference &   difference (\%) \\
\hline
  $(5s^2_{1/2}5p^2_{1/2}), J=0 $           &   3113  &   3153  &    40 &  1.3  \\
  $(5s^2_{1/2}5p^1_{1/2}5p^1_{3/2}), J=1 $ &   0     &   0     &   0   &   0   \\
  $(5s^2_{1/2}5p^1_{1/2}5p^1_{3/2}), J=2 $ &   5143  &   5139  &   -4  &  -0.1 \\
  $(5s^2_{1/2}5p^2_{3/2}), J=2 $           &   5941  &   5893  &   -48 &  -0.8 \\
  $(5s^2_{1/2}5p^2_{3/2}), J=0 $           &   15873 &   15820 &   -53 &  -0.3 
\end{tabular}

\begin{tabular} {lrrrr}
\hline
\hline
\multicolumn{5}{c}{  Lead atom} \\
\hline
\hline
  configuration, &    DC &   DCG &   absolute &    relative \\ 
  term
  &  &  &   difference &   difference (\%) \\
\hline
  $(6s^2_{1/2}6p^2_{1/2}), J=0 $           &     0   &   0      &   0     &     0   \\
  $(6s^2_{1/2}6p^1_{1/2}6p^1_{3/2}), J=1 $ &   4752  &   4644   &   -108  &    -2.3 \\
  $(6s^2_{1/2}6p^1_{1/2}6p^1_{3/2}), J=2 $ &   9625  &   9514   &   -111  &    -1.2 \\
  $(6s^2_{1/2}6p^2_{3/2}), J=2 $           &   18826 &   18592  &   -234  &    -1.2 \\
  $(6s^2_{1/2}6p^2_{3/2}), J=0 $           &   28239 &   27995  &   -244  &    -0.9   
\end{tabular}

\begin{tabular} {lrrrr}
\hline
\hline
\multicolumn{5}{c}{  Ekalead atom} \\
\hline
\hline
  configuration, &    DC &   DCG &   absolute &    relative \\ 
  term
  &  &  &   difference &   difference (\%) \\
\hline
  $(7s^2_{1/2}7p^2_{1/2}), J=0 $ &   0  &   0   &   0   &     0    \\
  $(7s^2_{1/2}7p^1_{1/2}7p^1_{3/2}), J=1 $ &   27198   &   26806    &   -392  &   -1.4  \\
  $(7s^2_{1/2}7p^1_{1/2}7p^1_{3/2}), J=2 $ &   30775   &   30391   &   -384  &    -1.2    \\
  $(7s^2_{1/2}7p^2_{3/2}), J=2 $ &   66068   &   65225  &   -843  &   -1.3   \\  
  $(7s^2_{1/2}7p^2_{3/2}), J=0 $ &   74527  &   73674  &   -853  &    -1.1    
\\
\hline
\hline
\end{tabular}
\label{Pb}
\end{table}

\begin{table}
\caption{Transition energies between different configurations of 
 the Indium($Z=49$), Thallium($Z=81$) and Ekathallium($Z=113$) atoms 
 obtained by Dirac-Hartree-Fock method with Coulomb and Coulomb-Gaunt 
 two-electron interactions.  All values except relative difference are 
 in $cm^{-1}$.}

\begin{tabular} {lrrrr}
\hline
\hline
\multicolumn{5}{c}{  Indium atom} \\
\hline
\hline
  configuration  &    DC &   DCG &   absolute &    relative \\ 
                     &  &  &   difference &   difference (\%) \\
\hline
  $(5s^2_{1/2}5p^1_{1/2}) $ &   0     &   0     &   0   &   0    \\
  $(5s^2_{1/2}5p^1_{3/2}) $ &   2142  &    2113 &   -29 &   -1.4 \\
  $(5s^2_{1/2}5d^1_{3/2}) $ &   30436 &   30377 &   -59 &   -0.2 \\ 
  $(5s^2_{1/2}5d^1_{5/2}) $ &   30453 &   30393 &   -60 &   -0.2 \\
  $(5s^2_{1/2}6s^1_{1/2}) $ &   22254 &   22201 &   -23 &   -0.2
\end{tabular}

\begin{tabular} {lrrrr}
\hline
\hline
\multicolumn{5}{c}{  Thallium atom} \\
\hline
\hline
  configuration  &    DC &   DCG &   absolute &    relative \\ 
                     &  &  &   difference &   difference (\%) \\
\hline
  $(6s^2_{1/2}6p^1_{1/2}) $ &   0     &    0    &   0    &   0    \\
  $(6s^2_{1/2}6p^1_{3/2}) $ &   7684  &    7597 &   -87  &   -1.1 \\
  $(6s^2_{1/2}6d^1_{3/2}) $ &   33035 &   32899 &   -136 &   -0.4 \\ 
  $(6s^2_{1/2}6d^1_{5/2}) $ &   33086 &   32949 &   -137 &   -0.4 \\
  $(6s^2_{1/2}7s^1_{1/2}) $ &   24141 &   24016 &   -125 &   -0.5
\end{tabular}

\begin{tabular} {lrrrr}
\hline
\hline
\multicolumn{5}{c}{  Ekathallium atom} \\
\hline
\hline
  configuration  &    DC &   DCG &   absolute &    relative \\ 
                     &  &  &   difference &   difference (\%) \\
\hline
  $(7s^2_{1/2}7p^1_{1/2}) $ &   0      &   0     &   0    &    0   \\
  $(7s^2_{1/2}7p^1_{3/2}) $ &   25447  &   25101 &   -346 &   -1.4 \\
  $(7s^2_{1/2}7d^1_{3/2}) $ &   45547  &   45152 &   -395 &   -0.9 \\ 
  $(7s^2_{1/2}7d^1_{5/2}) $ &   45654  &   45258 &   -396 &   -0.9 \\
  $(7s^2_{1/2}8s^1_{1/2}) $ &   35336  &   34962 &   -374 &   -1.1
\\
\hline
\hline
\end{tabular}

\label{Tl1}
\end{table}

\begin{table}
\caption{Hyperfine structure constants in Indium($Z=49$), Thallium($Z=81$)
 and Ekathallium($Z=113$) atoms calculated by Dirac-Hartree-Fock method with
 Coulomb and Coulomb-Gaunt two-electron interactions for different
 configurations. All values except relative difference are in MHz.}

\begin{tabular} {lrrrr}
\hline
\hline
\multicolumn{5}{c}{  Indium atom} \\
\hline
\hline
  configuration  &    DC &   DCG &   absolute &    relative \\ 
                     &  &  &   difference &   difference (\%) \\
\hline
  $(5s^2_{1/2}5p^1_{1/2}) $ &   1913   &   1900   &   -13    &    -0.7  \\
  $(5s^2_{1/2}5p^1_{3/2}) $ &   288    &   287    &   -1     &   -0.3   \\
  $(5s^2_{1/2}5d^1_{3/2}) $ &   4.41   &   4.40   &   -0.01  &   -0.2   \\ 
  $(5s^2_{1/2}5d^1_{5/2}) $ &   1.88   &   1.88   &    0.00  &    0.0   \\
  $(5s^2_{1/2}6s^1_{1/2}) $ &   1013   &   1011   &    -2    &   -0.2     
\end{tabular}

\begin{tabular} {lrrrr}
\hline
\hline
\multicolumn{5}{c}{  Thallium atom} \\
\hline
\hline
  configuration  &    DC &   DCG &   absolute &    relative \\ 
                     &  &  &   difference &   difference (\%) \\
\hline
  $(6s^2_{1/2}6p^1_{1/2}) $ &   18918   &   18691 &   -227  &   -1.2 \\
  $(6s^2_{1/2}6p^1_{3/2}) $ &   1403    &   1391  &   -12   &   -0.9 \\
  $(6s^2_{1/2}6d^1_{3/2}) $ &    20.8   &   20.8  &     0.0 &   0.0  \\ 
  $(6s^2_{1/2}6d^1_{5/2}) $ &    8.72   &   8.70  &   -0.02 &   -0.2 \\
  $(6s^2_{1/2}7s^1_{1/2}) $ &   7826    &   7807  &    -19  &   -0.2     
\end{tabular}

\begin{tabular} {lrrrr}
\hline
\hline
\multicolumn{5}{c}{  Ekathallium atom $^{\rm a}$} \\
\hline
\hline
  configuration  &    DC &   DCG &   absolute &    relative \\ 
                     &  &  &   difference &   difference (\%) \\
\hline
  $(7s^2_{1/2}7p^1_{1/2}) $ &   150168  &   147538  &    -2630 &   -1.8 \\
  $(7s^2_{1/2}7p^1_{3/2}) $ &   2007   &   1983     &    -24   &   -1.2 \\
  $(7s^2_{1/2}7d^1_{3/2}) $ &   34.3   &  34.2      &    -0.1  &   -0.3 \\ 
  $(7s^2_{1/2}7d^1_{5/2}) $ &   13.5   &   13.5     &     0.0  &    0.0 \\
  $(7s^2_{1/2}8s^1_{1/2}) $ &   28580   &   28473   &    -107  &   -0.4 \\
\hline
\end{tabular}

\noindent{  $^{\rm a}$ The magnetic moment $\mu_N$ and spin $I$ for 
the Eka-thallium nucleus were taken as those for Thallium.
The presented results can be easily recalculated as only the proper values
of $\mu_N$ and $I$ are known because they just include 
the $\mu_N/I$ coefficient.}
\label{Tl2}
\end{table}

\begin{table}
\begin{center}
\caption{
         Core-core plus core-valence contributions (i.e.\
         valence-valence contributions are excluded)
         to the total energy of the uranium atom 
         from the Gaunt and retardation interactions 
	 for different choices of the cores
         (in a.u.).
	 }
\vspace{5mm}
\begin{tabular}{crr|rr}
\hline
\hline
          &\multicolumn{2}{c}{ $5f^3 7s^2 6d^1$ } & \multicolumn{2}{c}{ $5f^2
7s^2 6d^2$ } \\
core      & Gaunt  &Retardation & Gaunt   & Retardation      \\
          & (DCG-DC) &(DCB-DCG)  & (DCG-DC)   & (DCB-DCG)  \\
\hline
 \protect[He\protect]                 & ~~~~28.62479082 & ~~~~-3.07749038 & ~~~~28.62643211 & ~~~~-3.07782300 \\
 \protect[Ne\protect]                 & 38.76297919     & -4.17703048     & 38.76744083     & -4.17791688 \\
\protect[Ar $3d^{10}$\protect]        & 41.53498862     & -4.40193727     & 41.53928838     & -4.40254968 \\
\protect[Kr $4d^{10}4f^{14}$\protect] & 42.00441512     & -4.42817590     & 42.00875116     & -4.42883266 \\
 all                                  & 42.02222003     & -4.43058931     & 42.02637715     & -4.43120398 \\
\hline
\hline
\end{tabular}
\label{contr1}
\end{center}
\end{table}

\begin{table}
\begin{center}
\caption{
 Core-core plus core-valence contributions to the energy of the transition
        $5f^3 7s^2 6d^1 \rightarrow 5f^2 7s^2 6d^2$ from the Gaunt and
	retardation interactions calculated from the data of
Table~\ref{contr1}
        (in \cm).}
\vspace{5mm}
\begin{tabular}{cccccc}
\hline
\hline
core               &   [He]   & [Ne]   & [Ar $3d^{10}$] 
&  [Kr $4d^{10}4f^{14}$]   &  all   \\
\hline
Gaunt              & 360   &  979  & 944    & 952   &  912   \\
(DCG-DC)           &       &       &        &       &        \\
Retardation        & -73   & -195  & -134   & -144  &  -135  \\
(DCB-DCG)          &       &       &        &       &        \\
\hline
\hline
\end{tabular}
\label{contr2}
\end{center}
\end{table}

\begin{table}
\begin{center}
\caption{
         The BI contributions to the total energy from different shells,
         calculated using \Eq{Bij0C} and estimated by \Eq{Bsim} (in \cm).
	 }
\vspace{5mm}
\begin{tabular}{ccc|c||}
\hline
\hline
\multicolumn{4}{c}{valence-valence contributions} \\
           &\multicolumn{2}{c}{ $ns^2$ } & $ns^2np^6$ \\
atom       &       $\rm{BI^a}$     & $\alpha^2\langle ns|\frac{1}{r}|ns\rangle^3 \cdot c_{N}^b$  & BI \\
\hline
He (n=1)   & 14 (14)       &   19   &                \\
Ne (n=2)   & 15 (21)       &   18   &  250 (488) \\
Ar (n=3)   &  3.4 (6.5)    &  3.6   &  55 (126)  \\
Kr (n=4)   & 2.2 (5.1)     &  2.2   &  36 (90)   \\
Xe (n=5)   & 1.3 (3.6)     &  1.3   &  23 (62)   \\
Rn (n=6)   & 1.2 (4.1)     &  1.2   &  23 (64)   \\
U  (n=7)   & 0.094 (0.27)  &  0.096 &                \\
\hline
\hline
\end{tabular}
\begin{tabular}{crr}
\hline
\hline
\multicolumn{3}{c}{core-core contributions} \\
           &\multicolumn{2}{c}{ $ns^2$ }  \\
atom       &       BI      & $\alpha^2\langle ns|\frac{1}{r}|ns \rangle ^3\cdot c_{N} $ \\
\hline
He (n=1)   &   14    &       19  \\
Ne (n=1)   & 2602    &     3563  \\
Ar (n=1)   & 15880   &    22045  \\
Kr (n=1)   & 133442  &   197531  \\
Xe (n=1)   & 468959  &   779730  \\
Rn (n=1)   & 2095503 & 5170727   \\
U (n=1)    & 2634137 & 7290932   \\
\hline
\hline
\end{tabular}
\vspace{5mm}

\begin{tabular}{ccc}
\hline
\hline
\multicolumn{3}{c}{core-valence contributions for uranium} \\
           &\multicolumn{2}{c}{ $ns^2ms^2$ }  \\
shells       &       BI      & $\alpha^2\langle ns|\frac{1}{r}|ns \rangle^2 \cdot
 \langle ms|\frac{1}{r}|ms \rangle \cdot c_{N}\cdot 2.7^c $ \\
\hline
(n=7, m=6)   &   .73  & .79   \\
(n=7, m=5)   &  1.9   & 1.9   \\
(n=7, m=4)   &  4.6   & 4.3   \\
(n=7, m=3)   & 11     & 10    \\
(n=7, m=2)   & 33     & 28    \\
(n=7, m=1)   & 151    & 109   \\
(n=6, m=5)   & 21     & 18    \\
(n=6, m=4)   & 51     & 41    \\
(n=6, m=3)   & 123    & 95    \\
(n=6, m=2)   & 360    & 260   \\
(n=6, m=1)   & 1658   & 1028  \\
\hline
\hline
\end{tabular}

\noindent (a) {The values obtained with eq.~(\ref{rad2})
               instead of eq.~(\ref{rad1}) are in brackets.}

\noindent (b) {The normalizing coefficient, $c_{N}\,{=}\,0.34$, is calculated
               as some average value for the BI contributions within the
               considered $s$-shells.}

\noindent (c) {The additional factor, 2.7, to $c_{N}$ appears due to the fact
               only the exchange ($K_B$) terms of BI between four electron
               pairs from different shells, $ns^2$ and $ms^2$ ($n \ne m$), are
               considered, whereas both the direct ($J_B$) and exchange
               ($K_B$) BI terms between the only one electron pair contribute
               in the case of the same shell, $ns^2$.  Taking into account that
                 $\left|K_B\right| \approx 2\cdot \left|J_B\right|$
               (this expression is exact in the case $n = m$), 
               one goes to the value $8/3 \approx 2.7$.}

\label{valence}
\end{center}
\end{table}

\bibliographystyle{bib/apsrev}
\bibliography{bib/JournAbbr,bib/TitovLib,bib/Titov,bib/Isaev}
\end{document}